\documentclass[a4paper]{jpconf}
\usepackage{graphicx}
\usepackage{iopams}

\begin{document}
\title{Contribution of incoherent effects to the orientation dependence of bremsstrahlung from rapid electrons in crystal}

\author{N F Shul'ga$^1$, V V Syshchenko$^2$ and A I Tarnovsky$^2$}

\address{$^1$ Akhiezer Institute for Theoretical Physics of the NSC
``KIPT'', Akademicheskaya Street, 1, Kharkov 61108, Ukraine}

\address{$^2$ Belgorod State University, Pobedy Street, 85, Belgorod 308015, Russian Federation}

\ead{shulga@kipt.kharkov.ua, syshch@bsu.edu.ru, syshch@yandex.ru}

\begin{abstract}
The bremsstrahlung cross section for relativistic electrons in a
crystal is split into the sum of coherent and incoherent parts
(the last is due to a thermal motion of atoms in the crystal).
Although the spectrum of incoherent radiation in crystal is
similar to one in amorphous medium, the incoherent radiation
intensity could demonstrate substantial dependence on the crystal
orientation due to the electrons' flux redistribution in the
crystal. In the present paper we apply our method of the
incoherent bremsstrahlung simulation developed earlier to
interpretation of some recent experimental results obtained at the
Mainz Microtron MAMI.
\end{abstract}

\section{Introduction}
It is well known (see, e.g. \cite{TM, AhSh, Ugg}) that high energy
electron beam incident on an oriented single crystal produces the
coherent radiation that is due to the spatial periodicity of the
lattice atoms, and the incoherent one, that is due to the thermal
spread of atoms from their positions of equilibrium in the
lattice. For the first look, the incoherent part of radiation is
similar to the last in amorphous medium (with Bethe-Heitler
spectrum), and do not depend on the crystal orientation in
relation to the particles beam.

However, in \cite{Sh2, Sh3} it was paid attention to the fact that
some features of the particle's dynamics in the crystal
(channeling effect etc.) could lead to various substantial
orientation effects in the hard range of the spectrum, where (for
$\varepsilon\sim 1$ GeV electrons) the incoherent part is
predominant. The semi-numerical approach developed in \cite{Sh2,
Sh3} was used for interpretation of early experimental data
\cite{Sanin}.

The ideass of \cite{Sh3} had been referred by the authors of
recent experiments \cite{Backe} to interpret some of their
results. In our article we present the results of simulation of
the incoherent radiation under the conditions of the experiment
\cite{Backe}. A good agreement with the experimental data confirms
the interpretation given in \cite{Backe}.

For the reader convenience, in the next section we outline some
theoretical ideas of our approach.

\section{Bremsstrahlung in dipole approximation}

Radiation of relativistic electron in matter develops in a large
spatial region along the particle's momentum. This region is known
as the coherence length (or formation length) \cite{TM, AhSh}
$l_{\mathrm{coh}} \sim 2\varepsilon\varepsilon ' /m^2c^3\omega$,
where $\varepsilon$ is the energy of the initial electron,
$\omega$ is the radiated photon frequency, $\varepsilon ' =
\varepsilon -\hbar\omega$, $m$ is the electron mass, $c$ is the
speed of light. In the large range of radiation frequencies the
coherence length could exceed the interatomic distances in
crystal:
\begin{equation}
l_{\mathrm{coh}} \gg a. \label{sst:eq1}
\end{equation}
In this case the effective constant of interaction of the electron
with the lattice atoms may be large in comparison with the unit,
so we could use the semiclassical description of the radiation
process. In the dipole approximation the spectral density of
bremsstrahlung is described by the formula \cite{AhSh}
\begin{equation}
\frac{dE}{d\omega} = \frac{e^2\omega}{2\pi c^4} \int_\delta^\infty
\frac{dq}{q^2}\left[ 1+
\frac{(\hbar\omega)^2}{2\varepsilon\varepsilon '} - 2
\frac{\delta}{q} \left( 1-\frac{\delta}{q} \right) \right] \left|
\mathbf W_q \right|^2, \label{sst:eq2}
\end{equation}
where $\delta = m^2c^3\omega /2\varepsilon\varepsilon ' \sim
l_{\mathrm{coh}}^{-1}$, and the value
\begin{equation}
\mathbf W_q = \int_{-\infty}^\infty \dot{\mathbf v}_\perp (t)
e^{icqt} dt \label{sst:eq3}
\end{equation}
is the Fourier component of the electron's acceleration in the
direction orthogonal to its initial velocity.

The main contribution to the integral in \eref{sst:eq2} is made by
the small values of $q$, $q\sim\delta$. On the other hand, the
characteristic distances on which the electron's acceleration in
the field of the atom in \eref{sst:eq3} would substantially
distinct from zero are equal by the order of magnitude to the
atomic radius $R$. The corresponding time intervals in which the
integrand in \eref{sst:eq3} is distinct from zero are $\Delta
t\sim R/c$. But under the condition (\ref{sst:eq1}) in the
frequency range of our interest $\delta\ll R^{-1}$. So we could
present $\mathbf W_q$ in the form \cite{Sh4}
\begin{equation}
\mathbf W_q = c\sum_n \boldsymbol\vartheta_n e^{icqt_n},
\label{sst:eq4}
\end{equation}
where $\boldsymbol\vartheta_n$ is the two-dimensional electron
scattering angle under collision with the $n$-th atom, $t_n$ is
the time moment of the collision.

Consider now the radiation of the electron incident onto the
crystal under small angle $\psi$ to one of its crystallographic
axes. It is known \cite{TM, AhSh} that averaging of the equation
for $\left| \mathbf W_q \right|^2$ over the thermal vibrations of
atoms in the lattice leads to the split of this value (and so the
radiation intensity) into the sum of two terms describing coherent
and incoherent effects in radiation:
\begin{eqnarray}
\left< \left| \mathbf W_q \right|^2 \right> = c^2\sum_{n,m}
e^{icq(t_n-t_m)} \left< \boldsymbol\vartheta (\boldsymbol\rho_n +
\mathbf u_n) \right> \left< \boldsymbol\vartheta
(\boldsymbol\rho_m + \mathbf u_m) \right> \\
+ c^2\sum_n \left\{ \left< \left( \boldsymbol\vartheta
(\boldsymbol\rho_n + \mathbf u_n) \right)^2 \right> - \left(
\left< \boldsymbol\vartheta (\boldsymbol\rho_n + \mathbf u_n)
\right> \right)^2 \right\}, \label{sst:eq5}
\end{eqnarray}
where $\boldsymbol\rho_n = \boldsymbol\rho (t_n) -
\boldsymbol\rho_n^0$ is the impact parameter of the collision with
the $n$-th atom in its equilibrium position $\boldsymbol\rho_n^0$,
$\boldsymbol\rho (t)$ is the trajectory of the electron in the
plane orthogonal to the crystallographic axis (which could be
obtained by numerical integration of the equation of motion), and
$\mathbf u_n$ is the thermal shift of the $n$-th atom from the
position of equilibrium. In the range of radiation frequencies for
which
\begin{equation}
l_{\mathrm{coh}} \ll a/\psi , \label{sst:eq6}
\end{equation}
where $a$ is the distance between two parallel atomic strings the
closest to each other, the incoherent term (\ref{sst:eq5}) makes
the main contribution into the bremsstrahlung intensity
(\ref{sst:eq2}).

The radiation by the uniform beam of particles is characterized by
the radiation efficiency, that is the radiation intensity
(\ref{sst:eq2}) integrated over impact parameters of the
particles' incidence onto the crystal in the limits of one
elementary cell. So, the efficiency is the classical analog of the
quantum cross section. In the further consideration we shall
compare the radiation efficiency in the crystal to the
Bethe-Heitler efficiency of bremsstrahlung in amorphous medium.

For further computational details see \cite{Sh2, Sh3}.

\section{Origin of the orientation dependence of the incoherent bremsstrahlung}
When charged particles are incident onto the crystal under small
angle $\theta$ to one of the atomic planes densely packed with
atoms, the channeling phenomenon could takes the place (see, e.g.,
\cite{AhSh, Ugg}). Under planar channeling the electron moves in
the potential well formed by the attractive continuum potential of
the atomic plane (see figure 1, left panel). The largest incidence
angle, for which the capture into the channel is possible, is
called as the critical channeling angle $\theta_c$ \cite{AhSh,
Ugg}.

\begin{figure}[h]
\includegraphics[scale=0.5]{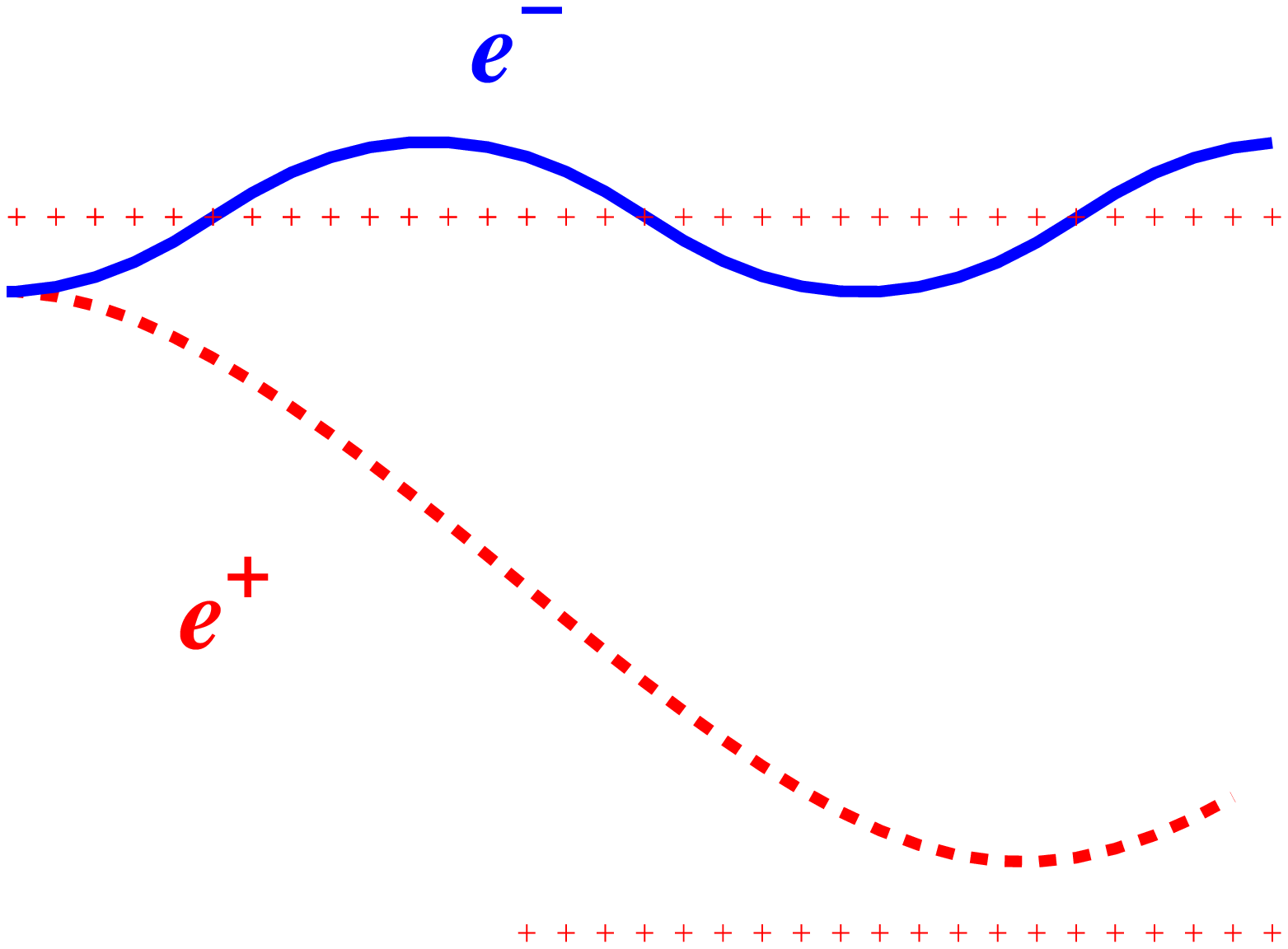} \ \
\includegraphics[scale=0.5]{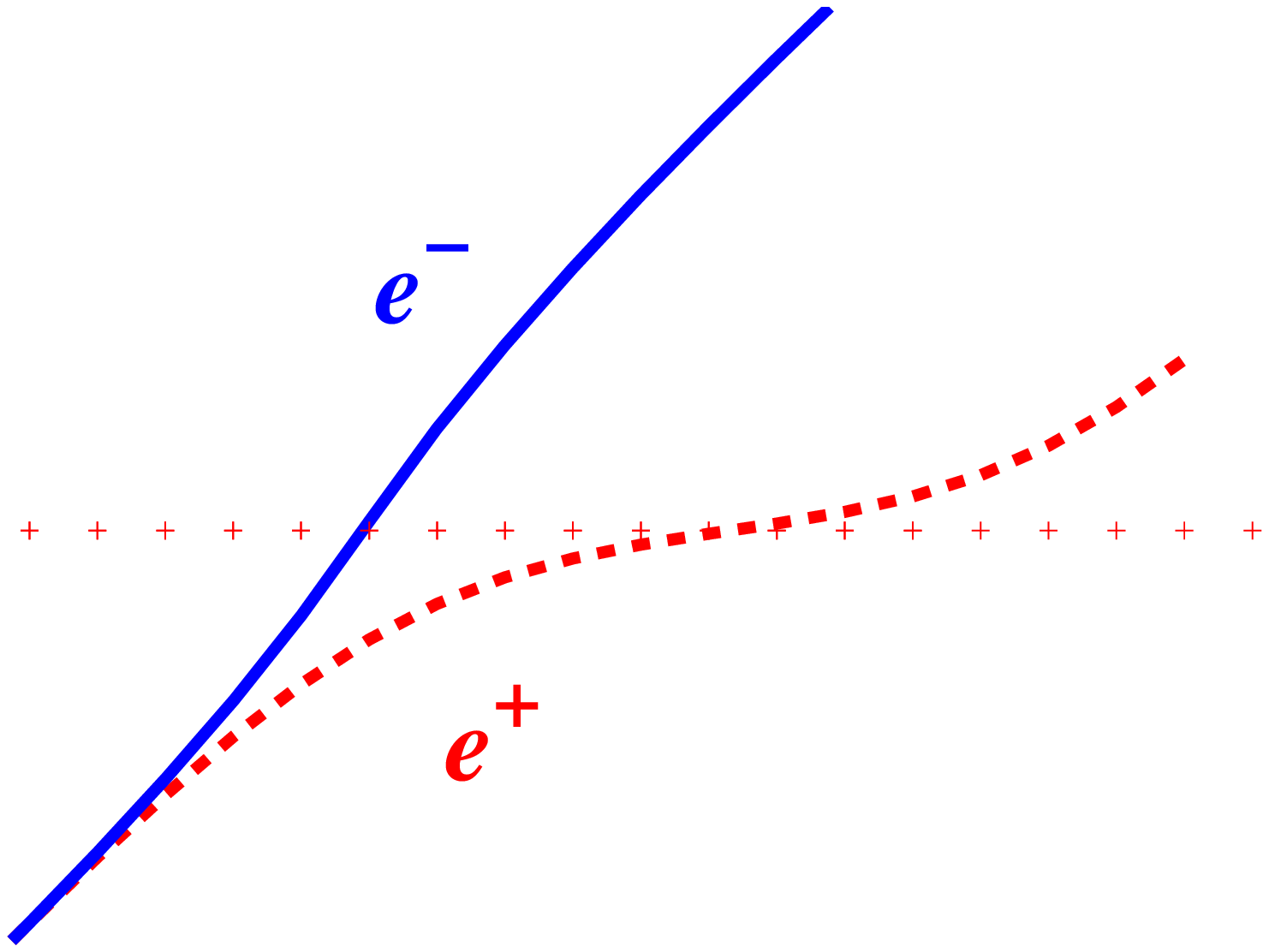}
\caption{Typical trajectories of the electrons (\full) and
positrons (\dashed) under planar channeling (left) and
above-barrier motion (right). Pluses mark the positions of atomic
strings (perpendicular to the plane of the figure) forming the
atomic planes of the crystal.}
\end{figure}

\begin{figure}[h]
\includegraphics[scale=0.55]{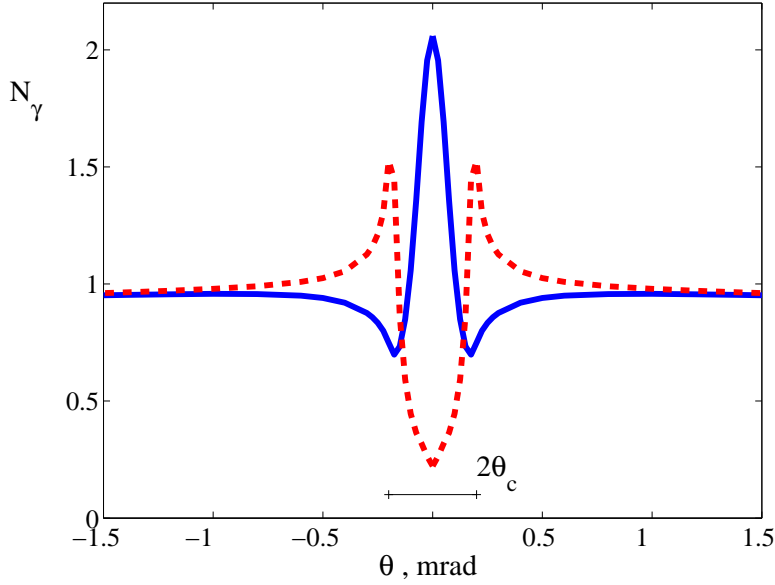}\hspace{1pc}
\begin{minipage}[b]{12pc}\caption{Incoherent bremsstrahlung efficiency (in ratio to the
Bethe-Heitler efficiency in amorphous medium) from 1 GeV electrons
(\full) and positrons (\dashed) vs incidence angle $\theta$ to
$(0\bar 11)$ plane of Si crystal, as a result of simulation.}
\end{minipage}
\end{figure}

Under $\theta\ll\theta_c$ the most part of the incident electrons
would move in the planar channeling regime. These electrons will
collide with atoms at small impact parameters more frequently then
in amorphous medium, that leads to the increase of the incoherent
bremsstrahlung efficiency (see figure 2). For $\theta\sim\theta_c$
the above-barrier motion in the continuum potential takes the
place for the most part of the particles (figure 1, right panel).
Above-barrier electrons rapidly cross the atomic plane, with
reduced number of close collisions with atoms comparing to the
case of amorphous medium. This leads to the decrease of the
incoherent bremsstrahlung efficiency (figure 2). For the positron
beam the situation is opposite.

\section{Results and discussion}
In one of the experiments of \cite{Backe} the integral intensity
of radiation (with the photon energy $\hbar\omega\geq 15$ MeV,
that eliminates the contribution of the channeling radiation) had
been measured for different orientations of the silicon crystals
in relation to 855 MeV electron beam. Note that under scanning of
the goniometer angle $\phi$ both the angle of incidence to the
$[100]$ axis and the angle to $(0 \bar 1 1)$ plane (and other
planes of the crystal) were changing. The zero angle between the
plane $(0 \bar 1 1)$ and the axis of the electron beam is achieved
near $\phi = 26$ mrad.

The experimental data (\cite{Backe}, Fig. 14(b)) and the results
of our simulation are presented on the figure 3.

\begin{figure}[h]
\includegraphics[scale=0.54]{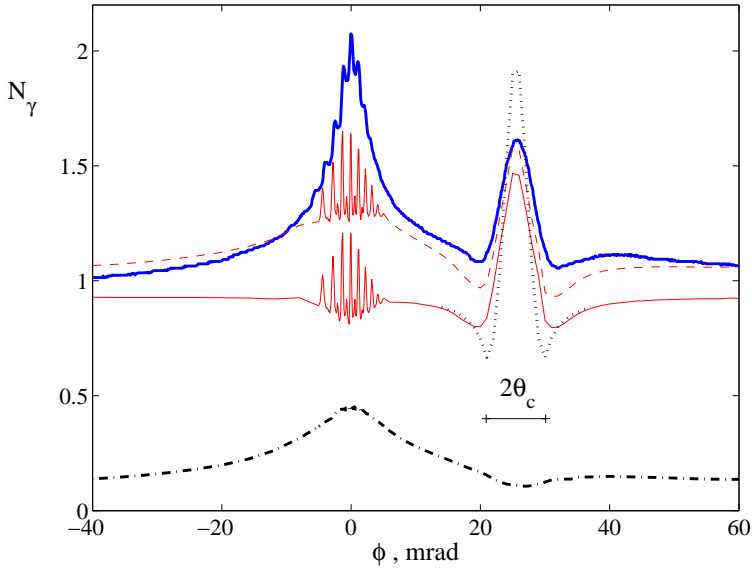}\hspace{1pc}
\begin{minipage}[b]{12.5pc}\caption{Coherent (\chain) and incoherent (\full) contributions to the relative
efficiency of radiation (\dashed) according to simulation, in
comparison to the experimental results \cite{Backe} (thick
curve).}
\end{minipage}
\end{figure}

The right peak, and the gaps surrounding it, are caused by the
contribution of incoherent radiation, as it was illustrated on the
figure 2. Dotted curve presents the result of simulation without
account of the incoherent multiple scattering of the electron on
the thermal vibrations of the lattice atoms; taking that into
account leads to some softening of the curve.

The left peak is caused by the effect of many crystallographic
planes with common $[100]$ axis. The fine structure of this peak
is also determined by the orientation dependence of the {\it
incoherent} radiation, as described above (see figure 4). The
crystallographic planes responsible to the origin of specific
local maxima in the incoherent radiation intensity are named on
the figure.

\begin{figure}[h]
\includegraphics[scale=0.54]{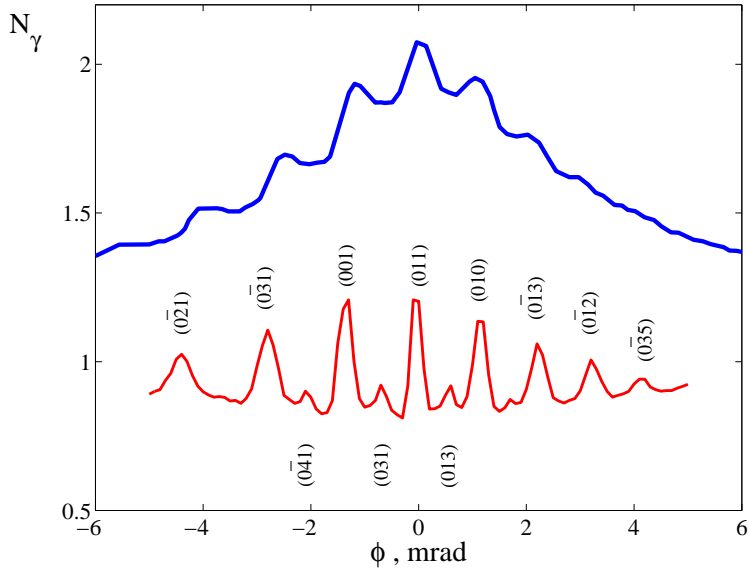}\hspace{1pc}
\begin{minipage}[b]{12.5pc}\caption{Region near $\phi=0$ of the previous figure.
Only experimental data (upper curve) and incoherent contribution
(lower curve) are shown.}
\end{minipage}
\end{figure}

The coherent contribution was calculated using standard Born
theory of the coherent bremsstrahlung \cite{AhSh}. We normalized
the experimental curve to the right edge of the calculated one,
far from the conditions of planar channeling, where Born theory of
the coherent bremsstrahlung is surely valid. For better agreement
with the experiment in the range near the left peak, it seems the
most suitable to carry out the simulation of the coherent
bremsstrahlung based on the same semiclassical approach, as used
for the study of the incoherent one \cite{Sh4}.

\ack This work is supported in part by the internal grant of
Belgorod State University.

\end{document}